\documentclass[11pt,a4paper]{scrartcl}

\usepackage{CLICdp}

\usepackage{CLICdp_definitions}

\newcommand{\ttb}{$\PQt \PAQt$\xspace}
\newcommand{\tch}{$\PQt \rightarrow \PQc \PH$\xspace}
\newcommand{\tcg}{$\PQt \rightarrow \PQc \PGg$\xspace}
\newcommand{\hbb}{$\PH \rightarrow \PQb \PAQb$\xspace}
\newcommand{\mtch}{\PQt \rightarrow \PQc \PH}
\newcommand{\mtcg}{\PQt \rightarrow \PQc \PGg}
\newcommand{\mhbb}{\PH \rightarrow \PQb \PAQb}
\newcommand{\lcfiplus}{\textsc{LcfiPlus}\xspace}

\title{Limits on top FCNC decay \tch and \tcg \\ from CLIC at 380 GeV}

\clicdpconf{2018}{001}  

\date{\today}

\addauthor{A.F.~\.Zarnecki\thanks{Filip.Zarnecki@fuw.edu.pl}}{\institute{1}}
\addauthor{N. van der Kolk\thanks{now at Nikhef, Amsterdam, the Netherlands}}{\institute{2}}
  
\addinstitute{1}{Faculty of Physics, University of Warsaw, Poland}
\addinstitute{2}{Max-Planck-Institut f\"ur Physik, Munich, Germany}

\onbehalfof{the CLICdp Collaboration} 

\abstract{FCNC top decays are very strongly suppressed in the Standard
  Model and the observation of any such decay would be a direct signature of
  physics beyond SM. Many "new physics" scenarios predict 
  contributions to FCNC processes and the largest enhancement in
  many models is for \tch decay. Enhancements for the decay
  channel \tcg are more modest, but the decay still has a clearly
  identifiable kinematic signature. Prospects for measuring these
  decays at CLIC running at 380 GeV were studied with full detector
  simulation, taking the luminosity distribution, beam polarization and
  beam induced background into account. Top pair production events
  with \tch decays can be identified based on the kinematic
  constraints and flavour tagging information. The analysis was divided
  into three steps: classification of top pair candidate events, event
  quality determination and kinematic reconstruction based on signal
  or background hypotheses, and final separation of signal from background.
  To obtain optimal results, selection criteria based
  on the dedicated Boosted Decision Trees (BDT) were used at each
  step. The expected limit on \mbox{BR(\tch)}$\times$\mbox{BR(\hbb)} from a 
  combined analysis of hadronic and semi-leptonic top pair samples, as
  well as the limit on BR(\tcg) from hadronic top pair decays are presented.
}

\titlecomment{Talk presented at the International Workshop on Future Linear Colliders
               (LCWS2017), Strasbourg, France, 23-27 October 2017. C17-10-23.2.}

\graphicspath{ {./plots/} }

\begin{document}


\titlepage


%
%

\section{Introduction}

Top physics, together with Higgs boson studies and searches for Beyond
the Standard Model (BSM) phenomena, is one of the three pillars of the
research programme for future high energy $\epem$ colliders.
As the top quark is the heaviest known elementary particle, with an
expected value of the Yukawa coupling of the order of one, the precise
determination of its properties is a key to understanding
electroweak symmetry breaking. 
The determination of top properties is also essential for many ``new
physics'' searches, as the top quark gives large loop contributions to many
precision measurements that are sensitive to BSM effects.
Stringent constraints on the new physics scenarios are also
expected from direct searches for rare top decays. 
Both future linear colliders, the International Linear Collider (ILC)
and the Compact Linear Collider (CLIC), provide an opportunity to
study the top quark with unprecedented precision via direct production
of thousands of $\PQt\PAQt$ pairs in $\epem$ collisions.

This contribution presents the prospects of constraining
the branching ratio for the flavour changing top decays \tcg and 
\tch with CLIC running
at $\sqrt{s} = 380$~GeV \cite{staging_baseline_yellow_report}.
These decays are very strongly suppressed in the Standard Model with the
expected branching ratios~\cite{Agashe:2013hma}:
\begin{eqnarray*}
\textrm{BR}(\mtcg) &  \approx & 5 \cdot 10^{-14}, \\
\textrm{BR}(\mtch) &  \approx & 3 \cdot 10^{-15}. 
\end{eqnarray*}
At the same time a significant enhancement of these decays is
expected in many extensions of the Standard Model, resulting from
either the introduction of the direct tree level FCNC couplings or
being due to the loop level 
contributions of new particles (or SM particles with modified couplings).
For the  \tcg decay an enhancement up to the level
of $10^{-5}$ is expected for some SUSY models with $R$-parity
violation~\cite{Mele:1999zx},
while for the \tch channel loop induced branching ratio of up to $10^{-4}$ is
predicted in many models~\cite{Bejar:2001sj}, with enhancement of up to
$10^{-2}$ possible on the tree level~\cite{DiazCruz:2006qy}.
Existing LHC constraints on the considered FCNC decays of the top quark are
rather weak, at the level of $2 \cdot 10^{-3}$
\cite{Aaboud:2017mfd,Khachatryan:2016atv,Khachatryan:2015att}
and the expected limits from HL-LHC are of the order of
$2 \cdot 10^{-4}$ \cite{Agashe:2013hma,ATL-PHYS-PUB-2016-019}. 
Measurements at CLIC can be competitive for these channels thanks
to the large sample of produced top quarks, clean environment and well
constrained kinematics.
The observation of any such decay would be a direct signature
for ``new physics''.

\section{Event simulation and reconstruction}

Detailed detector level analysis was performed for  $\epem$ collisions
at CLIC at $\sqrt{s} = 380$~GeV.
With the assumed lumionosity of 500~fb$^{-1}$ and electron beam
polarisation of -80\%  about 400~000 top quark pairs are expected.
Dedicated signal samples were generated with
WHIZARD 2.2.8 \cite{Kilian:2007gr,Moretti:2001zz}. 
For the \tcg channel the model with anomalous top couplings
(\verb|SM_top_anom|) with vector type tensor $\PQt\PQc\PGg$ coupling
was used while for the \tch channel the simulation was based on the
2HDM(III) model implemented in SARAH~\cite{Staub:2015kfa}.
For both models the parameter values were tuned to obtain a FCNC
branching ratio of $10^{-3}$ so that the contribution from events with two
FCNC decays is negligible. 
Detailed beam spectra for CLIC as well as beam induced backgrounds 
and the assumed electron beam polarization were taken into account.
Generated events were passed to PYTHIA~6.4 for
hadronisation with quark masses and other settings adjusted to the
configuration used previously in CLIC CDR studies~\cite{CLIC_PhysDet_CDR}.
Signal samples were then processed with a standard event
simulation and reconstruction chain of the CLICdp collaboration using
the CLIC\_ILD\_CDR500 detector configuration. 
The background sample considered in the presented analysis included a
full set of six-fermion event samples originally produced for CLICdp
studies of top pair production at $\sqrt{s} = 380$~GeV.
All sub-samples corresponded to an integrated luminosity of at least
500~fb$^{-1}$. 

Analysis of the two top decay channels presented in this contribution was
based on the so-called particle flow reconstruction performed using
\pandora \cite{thomson:pandora, Marshall2013153, Marshall:2015rfa}.
The reconstructed object collection resulting from loose background
rejection cuts \cite{CLIC_PhysDet_CDR} was used as an input
for jet reconstruction with the Valencia
algorithm~\cite{Boronat:2016tgd} as well as for primary and secondary 
vertex reconstruction, and flavour tagging with \lcfiplus
\cite{Suehara:2015ura}. 


\section{\tcg analysis}

\begin{figure}
\begin{center}
\includegraphics[width=0.6\textwidth]{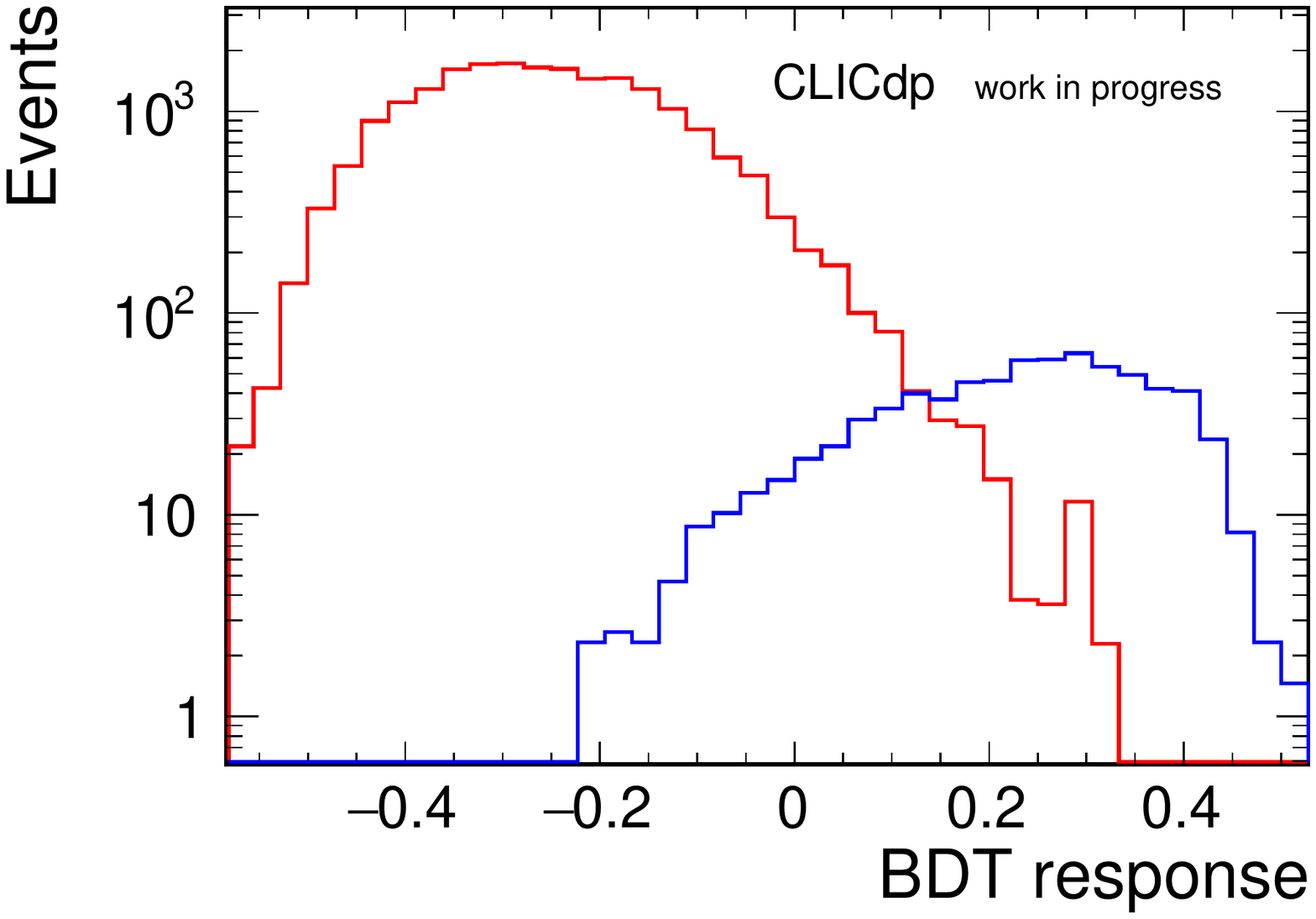}
\end{center}
\caption{Distribution of the BDT classifier response for events with 
    FCNC top decay  \tcg (signal, blue histogram)
    and SM top pair events (background, red histogram), for FCNC selection at 380 GeV
    CLIC. The background sample is normalised to 500~fb$^{-1}$ and the
    assumed signal level corresponds to  BR(\tcg) $=10^{-3}$.
\label{fig:tcg_bdt} }
\end{figure}

\begin{figure}
\begin{center}
\includegraphics[width=0.6\textwidth]{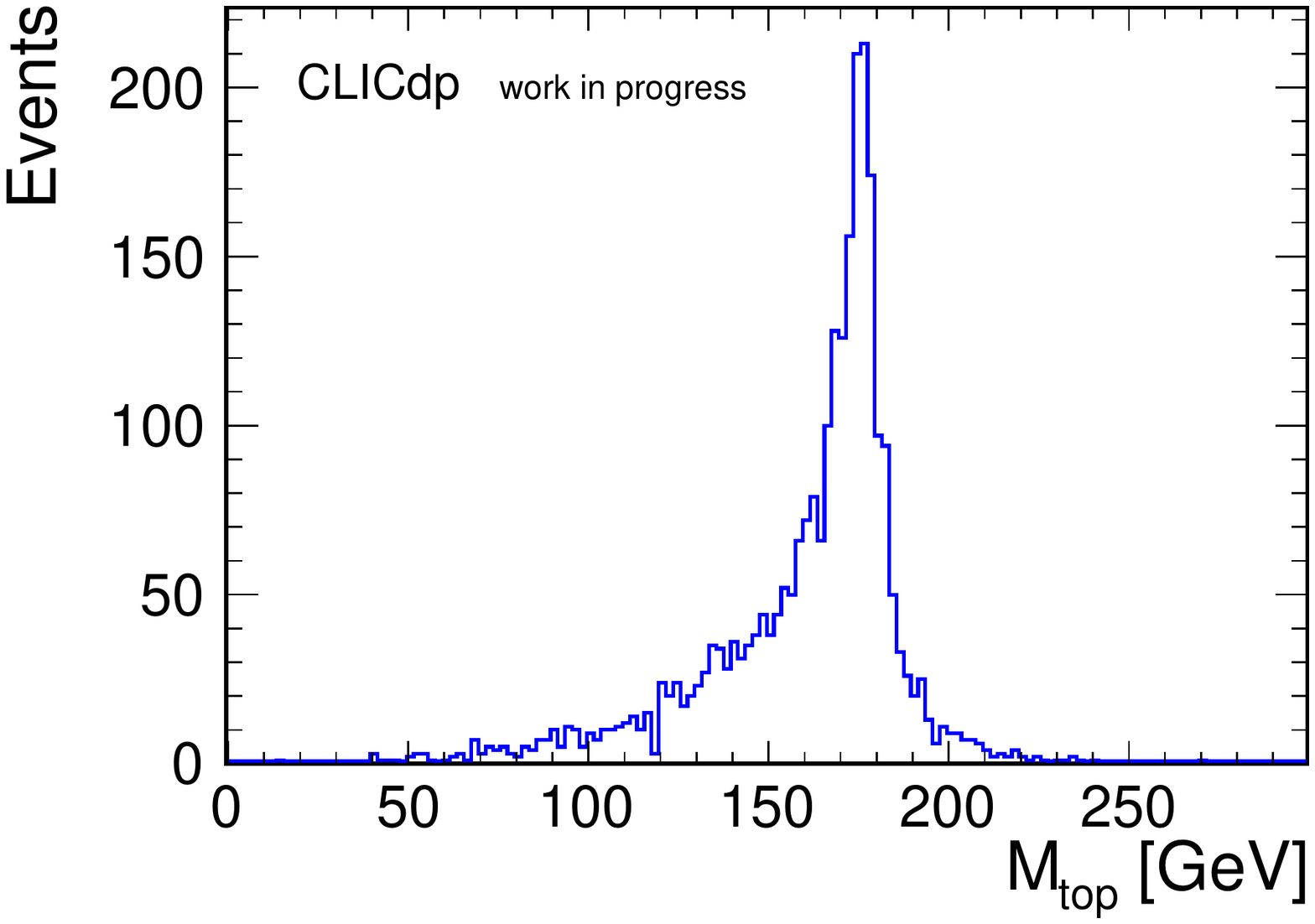}
\end{center}
\caption{Invariant mass distribution of the top quark from the
    FCNC decay \tcg reconstructed at 380~GeV CLIC after final
    selection cuts, for an integrated luminosity of 500~fb$^{-1}$ and
    assuming BR(\tcg) $=10^{-3}$.
\label{fig:tcg_mtop} }
\end{figure}

In this channel we search for top pair production events in which one of
the top quarks decays into a charm quark and a photon.
Due to the large top mass this decay results in a very striking
signature:  a high energy photon, with an energy of at least 50 GeV. 
Only the fully hadronic decay channel is considered where the
second (spectator) top quark decays into a b-quark and a W-boson, which
decays hadronically into two light quarks.
The analysis requires that an isolated photon with an energy of at
least 50 GeV is reconstructed in each selected event.
This simple cut reduces the background from standard \ttb decays
by a factor of 20 while keeping 92\% of the signal events.
Selected events are then subjected to kinematic fits, based on the jets
reconstructed using the Valencia jet clustering algorithm, for both the
signal  (\PGg + 4 jets) and background (6 jet) hypotheses.
The final selection of signal events is then based on a Boosted
Decision Tree (BDT) algorithm, as
implemented in the TMVA framework \cite{TMVA:2010}.
A total of 42 input variables were used to train the BDT, including photon
properties, jet properties, flavour tagging information, invariant
masses and $\chi^2$ values from the kinematic fits.
The distribution of the BDT classifier response for the considered
signal (FCNC decay events) and background (SM top decays) samples is
shown in Fig.~\ref{fig:tcg_bdt}.
%
%
Shown in Fig.~\ref{fig:tcg_mtop} is the reconstructed invariant mass
distribution for the signal top quark, after the final selection cut based
on the BDT response, BDT$>$0.2.
%
%
%
With this cut  60\% of the signal events are selected while reducing the
background contamination in the sample by a factor of about 500.
The total selection efficiency resulting from the initial selection and
BDT based selection is 55\% for signal events and $9.4 \cdot 10^{-5}$
for $\PQt\PAQt$ background. 
Assuming the nominal integrated luminosity of 500~fb$^{-1}$ collected at
380~GeV CLIC, the expected 95\% C.L. limit on the top FCNC decay
(assuming no signal contribution) is
\begin{eqnarray*}
   \textrm{BR}(\mtcg)  & < & 3 \cdot 10^{-5} \; . 
  \end{eqnarray*}


\section{\tch analysis}

The sensitivity of CLIC at 380 GeV to FCNC decay \tch was studied for the
dominant Higgs boson decay channel \hbb.
Six jets are expected in the final state for top pair production
events with one FCNC top quark decay and a hadronic decay of
the second (spectator) top quark, while four jets, an isolated lepton
and missing momentum are expected for events having a leptonic spectator decay.
These final states have topologies that are compatible with the
Standard Model top pair production events, which is therefore the dominant
background contribution for this analysis.
Discrimination between FCNC signal and SM top pair production events
has to be based on the kinematic fit results and the flavour
tagging information.
For signal events we expect three jets in the final state to be tagged
as b-quark jets (two from the Higgs boson and one from the
spectator top quark decay) with two of them consistent with the invariant
mass of the Higgs boson.
A cut based approach, which was considered at the initial stage of the
analysis \cite{Zarnecki:2017cmf}, was found to be too inefficient
for the considered process, as many observables had to be considered
at the same time.
This contribution presents the new analysis approach based on
the use of multivariate analysis.
BDT classifiers were used in the three analysis stages:
classification of top pair candidate events (into hadronic,
semi-leptonic and leptonic samples), estimation of the event reconstruction
quality and the final signal-background discrimination.

\begin{figure}
\begin{center}
\includegraphics[width=0.49\textwidth]{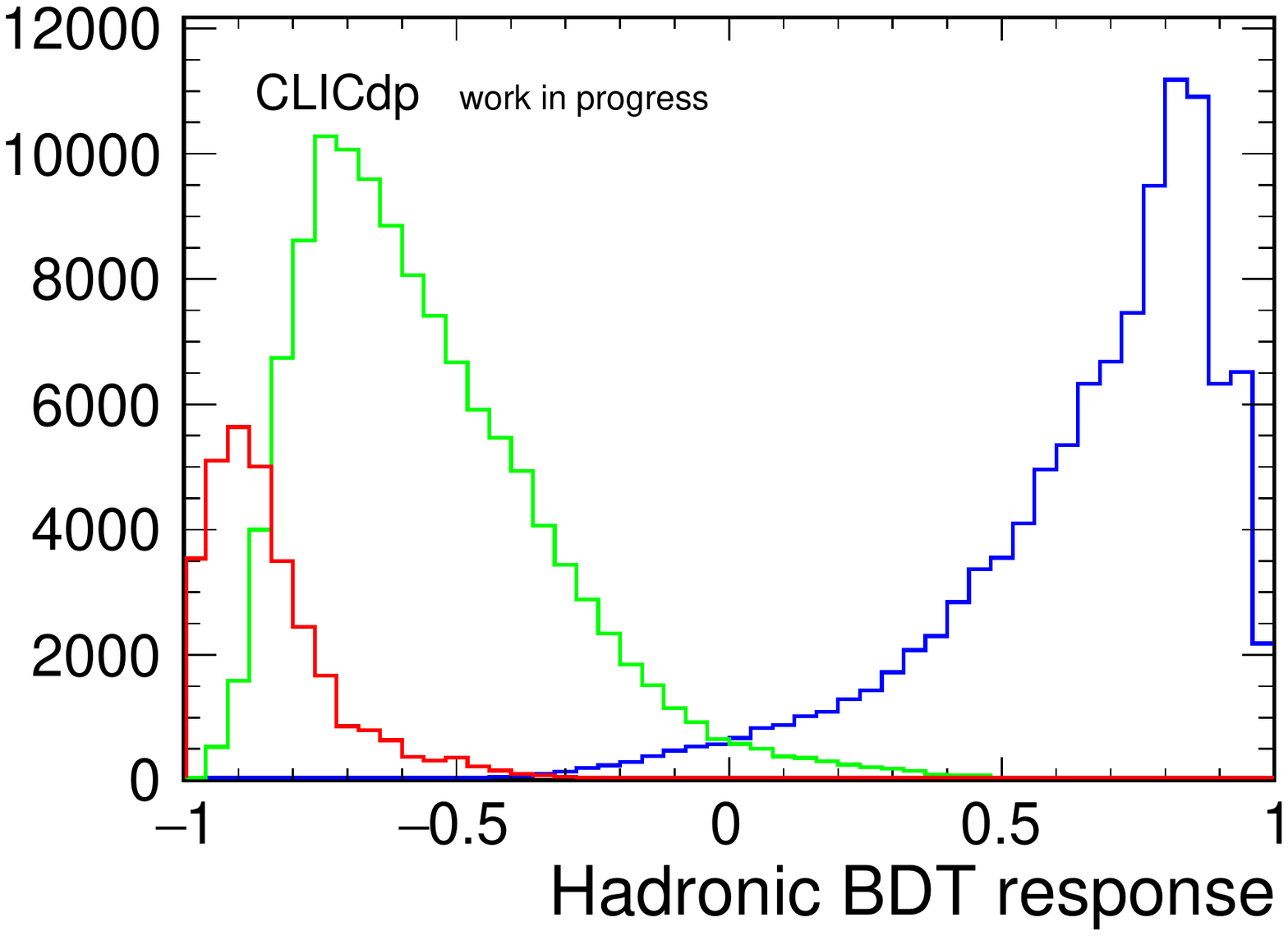}
\includegraphics[width=0.49\textwidth]{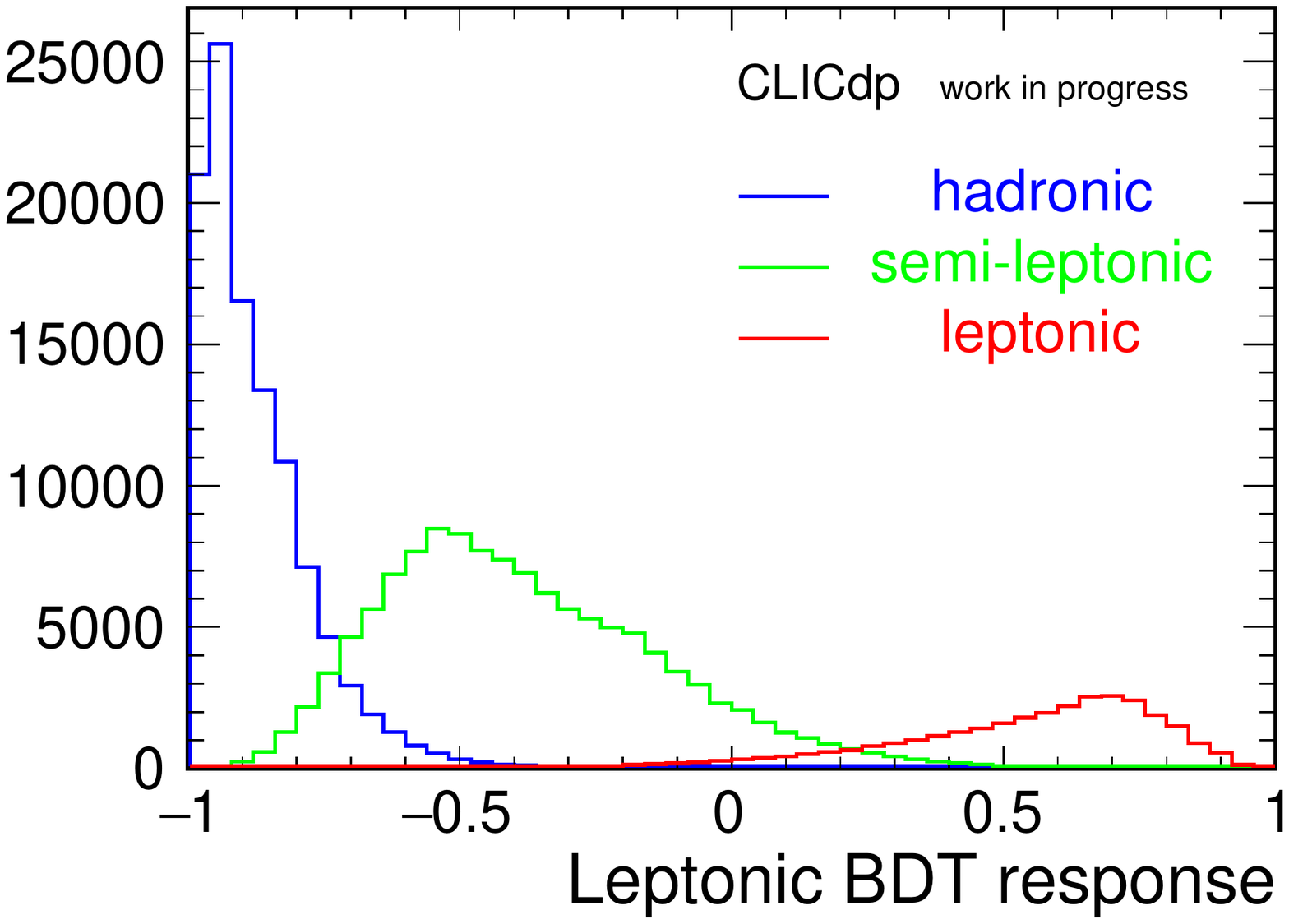}
\end{center}
\caption{Response distributions for BDT classifiers trained to recognize
    hadronic top pair events (left) and leptonic top pair events
    (right), for different samples of \ttb SM background events,
    for 500~ fb$^{-1}$ at 380 GeV CLIC. 
\label{fig:tch_hadlep}   }
\end{figure}

For selection of hadronic and semi-leptonic event samples, two
independent BDT classifiers, based on total event
energy-momentum, event shape variables, isolated lepton 
information and jet reconstruction parameters were prepared. 
Only SM background samples were used for BDT training at this stage,
with hadronic and purely leptonic top pair events used as a signal
when training the hadronic and leptonic BDT classifiers respectively.
Shown in Fig.~\ref{fig:tch_hadlep} are the response distributions for
these classifiers for different samples of top pair events.
%
%
A cut on the hadronic BDT response was used to select the hadronic event
sample, while a cut on both the hadronic and leptonic BDT responses was
required for semi-leptonic event selection.
Identification of the final state isolated lepton (electron or muon)
was also required for the semi-leptonic events.
The pre-selection based on the event classification included also the
initial (loose) cut on the flavour tagging results: three jets were
required to have a b-tag value of at least 0.4 and the fourth jet (jet
coming from the \PQc quark from FCNC decay) should have a sum of b-tag
and c-tag values of at least 0.4.

\begin{figure}
\begin{center}
\includegraphics[width=0.49\textwidth]{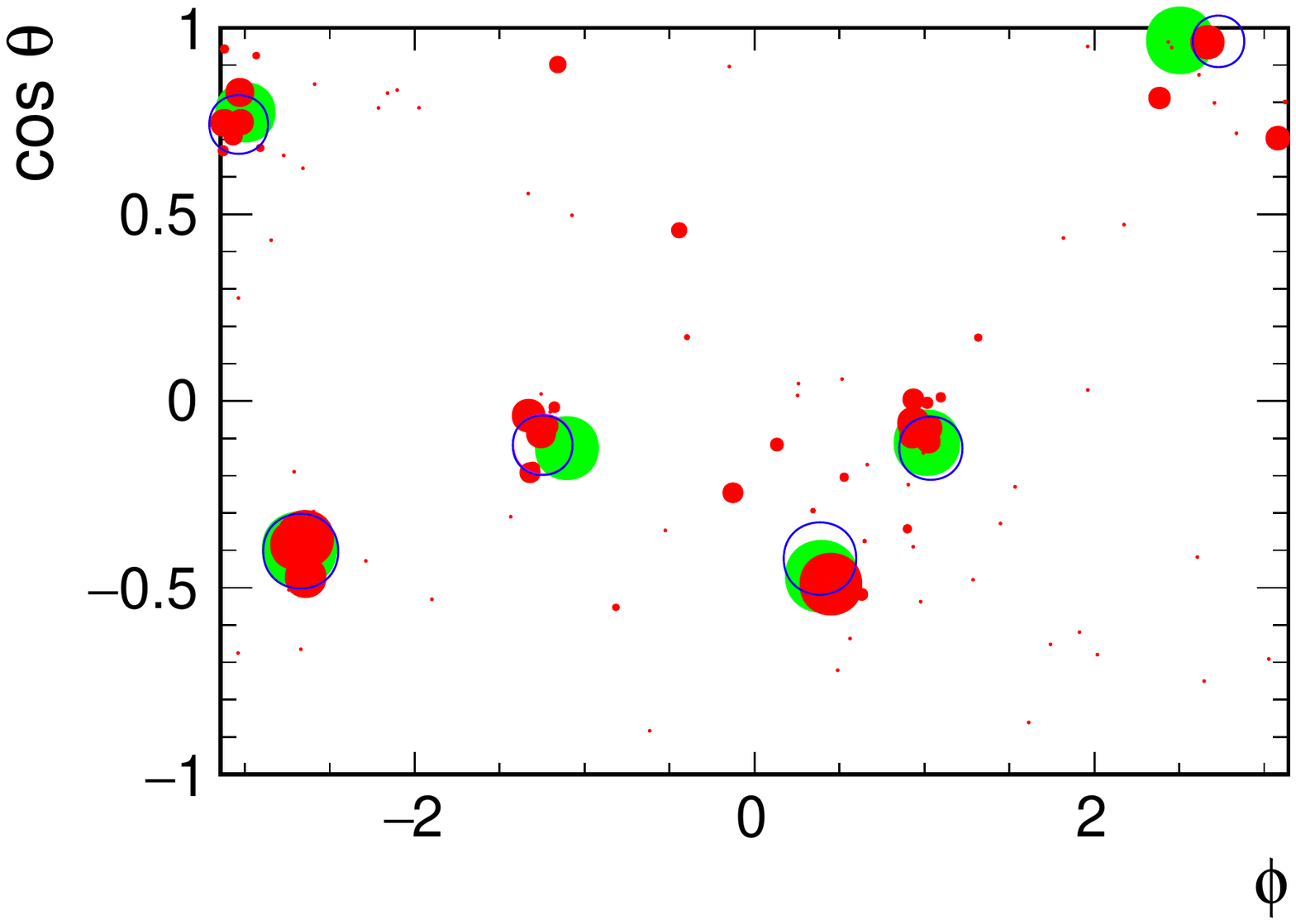}
\includegraphics[width=0.49\textwidth]{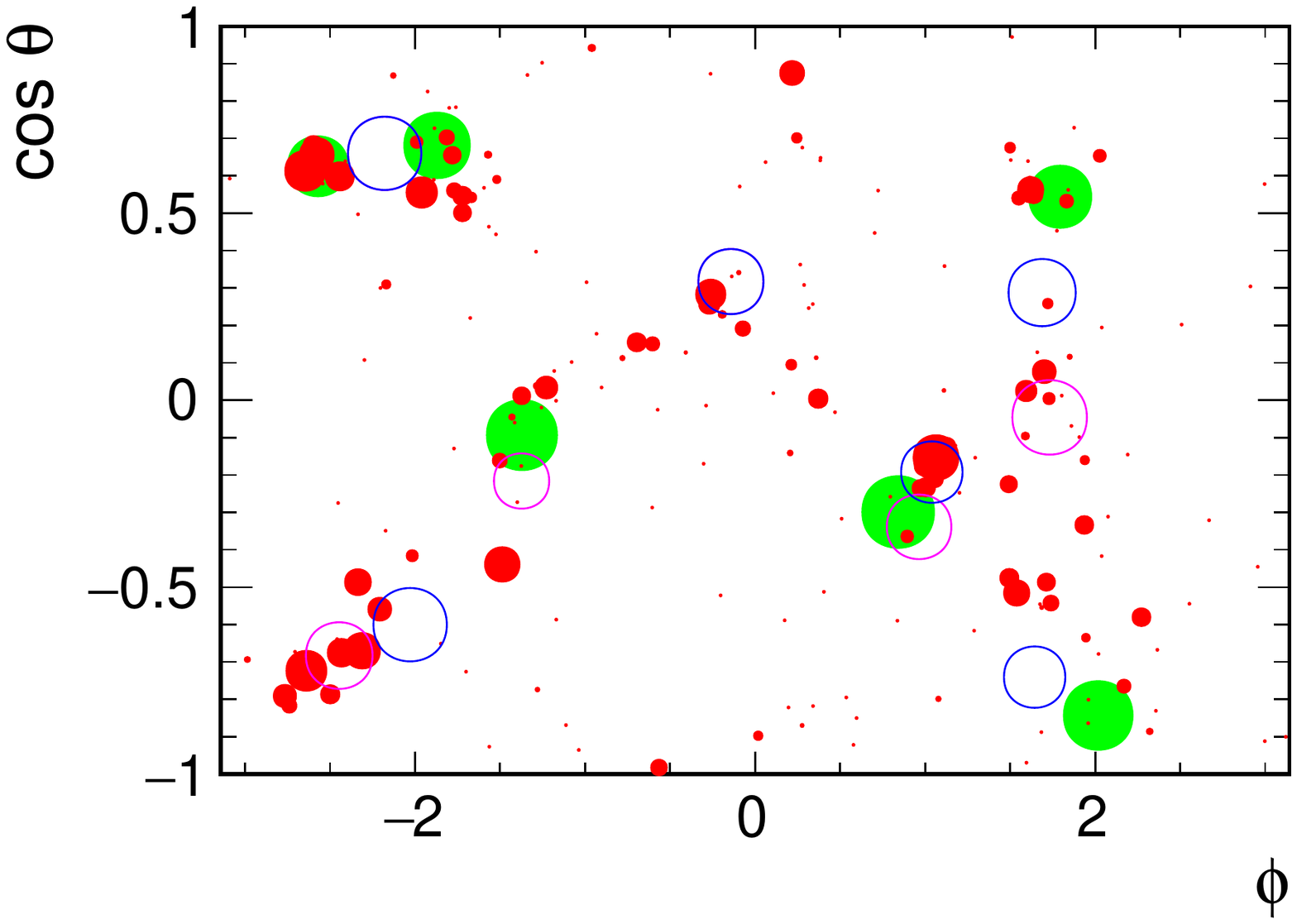}
\end{center}
\caption{Schematic view of two SM top pair production
    events with fully hadronic final state.
    Directions in ($\cos\theta$, $\phi$) of final state quarks
    (solid green circles) and reconstructed particle flow objects
    (solid red circles) are compared with clustering results of Valencia (open
    magenta circles) and anti-k$_{T}$ (open blue circles) jet
    algorithms.
\label{fig:tch_goodbad}   }
\end{figure}

As mentioned above, the signal-background discrimination for \tch channel
has to be based on the kinematic fit results.
However, the quality of the kinematic fit turned out to be very poor
for a significant fraction of events, both for the signal and for the
SM $\PQt\PAQt$ sample.
Figure~\ref{fig:tch_goodbad} shows two example top pair
production events from the hadronic sample.
Compared are the generator level directions of the final state quarks
with the reconstructed particle flow objects and clustering results of
Valencia algorithm (as used in the kinematic fit) and the anti-k$_{T}$ jet
clustering algorithms.
For the majority of the events (as the one shown in Fig.~\ref{fig:tch_goodbad}
left), reconstructed jets correspond closely to the partonic final state on
the generator level.
For such well-reconstructed events a kinematic fit allows for
precise determination of event kinematics and can be used for efficient
discrimination between FCNC and SM top quark decays. 
However, for a significant fraction of events (like the one shown in
Fig.~\ref{fig:tch_goodbad} right) final state particles and the
reconstructed particle flow objects do not follow the initial quark
directions.
%
%
This is mainly due to higher order QCD effects.
The quality of the kinematic fit is very poor for such poorly
reconstructed events and it cannot be used for efficient signal event
selection. 
It is therefore important to try to discriminate between well
reconstructed and poorly reconstructed events.

\begin{figure}
\begin{center}
\includegraphics[width=0.49\textwidth]{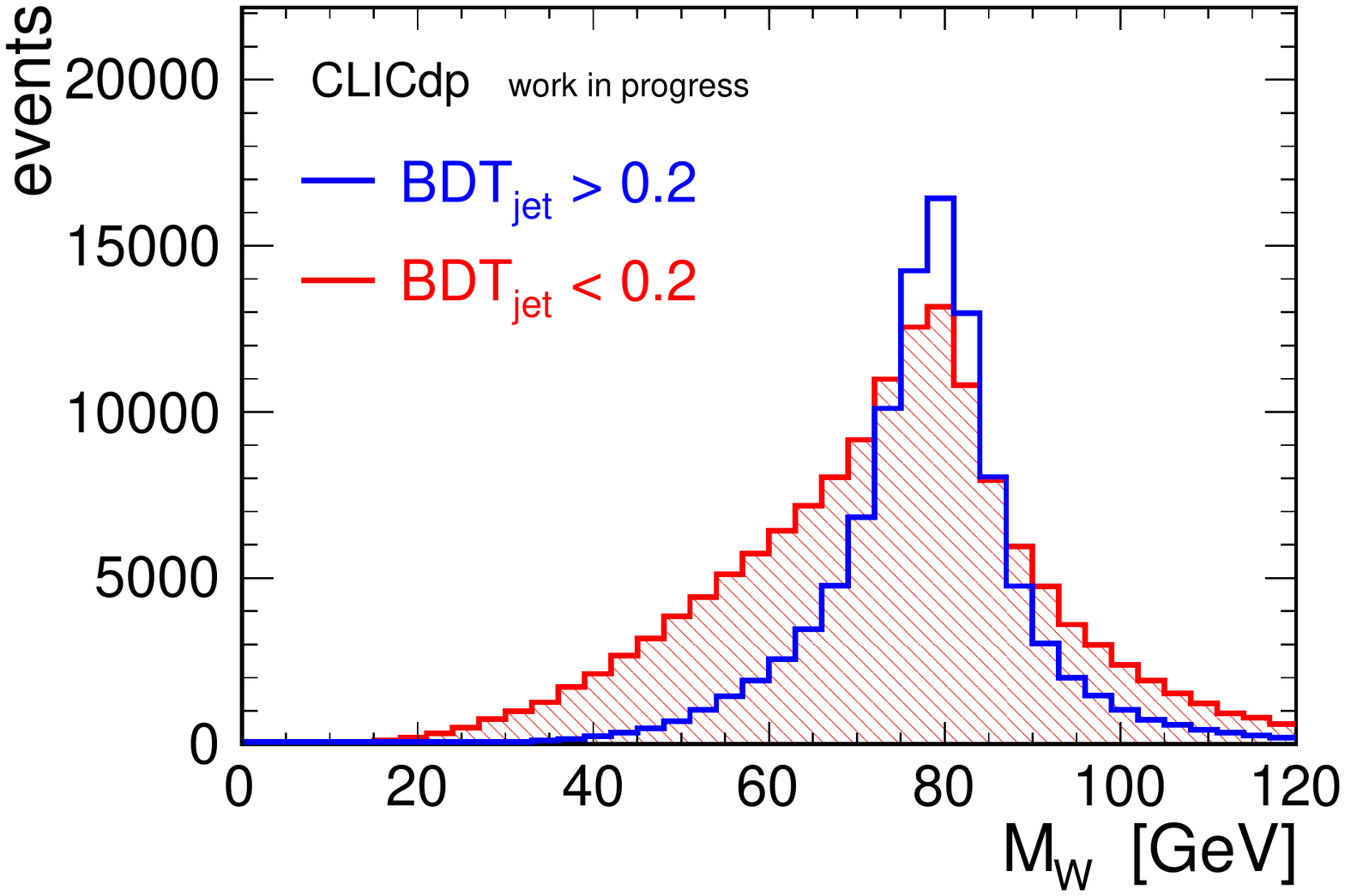}
\includegraphics[width=0.49\textwidth]{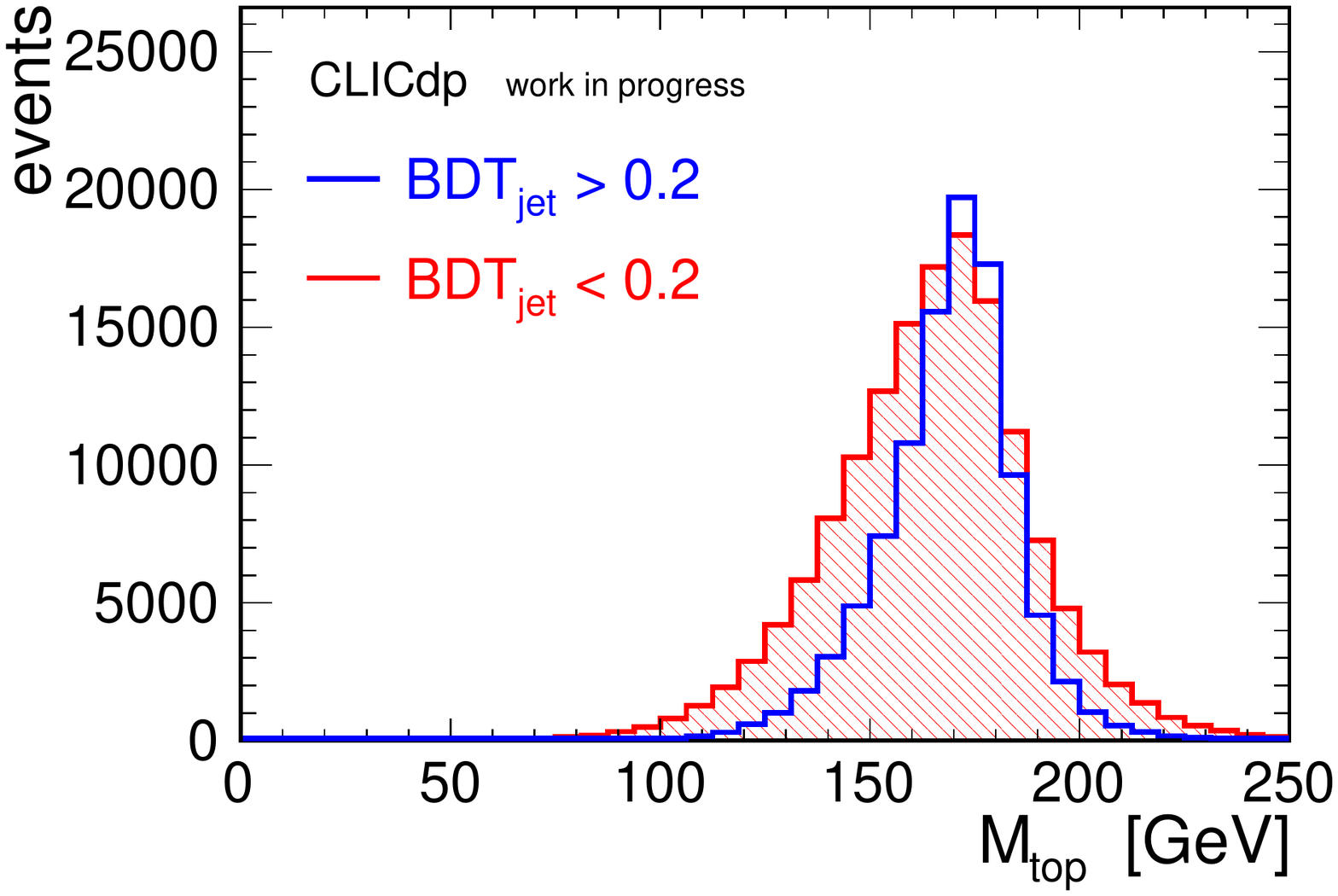}
\end{center}
\caption{Influence of the cut on the event quality classifier
    response, BDT$_{jet}$, on the kinematic reconstruction of hadronic events.
    Reconstructed invariant masses of the W bosons (left)
     and top quarks (right) for SM top pair events (normalised to  500~fb$^{-1}$).
\label{fig:tch_minv}   }
\end{figure}

The results of the kinematic fits should not be used to make any
selection at this stage, as they will be used in the final step to
discriminate between signal (\tch) and background ($\PQt \rightarrow
\PQb\PWp$) hypotheses. 
However, one can notice that for poorly reconstructed events the clustering
results have a much stronger dependence on the clustering algorithm applied.
It is therefore possible to estimate the event reconstruction quality
by comparing results of different clustering algorithms.
The BDT algorithm was trained separately for hadronic and
semi-leptonic events using the angular distances and energy
differences between jets reconstructed with three different jet
algorithm configurations.
The angular distance between the Valencia jets and the six-fermion
final state on the generator level was used to define the reference
samples for BDT training.
The obtained event quality estimate (BDT response, BDT$_{jet}$) is
only weakly correlated with the true parton-jet matching, but it does
improve significantly the quality of kinematic reconstruction.
Shown in Fig.~\ref{fig:tch_minv} is the influence of the cut on the
quality estimate on the reconstructed invariant mass of the \PWpm
boson and the top quark, for the sample of hadronic $\PQt\PAQt$ events.
The cut applied on the quality estimate significantly reduces
the tails of the invariant mass distributions.
%
%

\begin{figure}
\begin{center}
\includegraphics[width=0.49\textwidth]{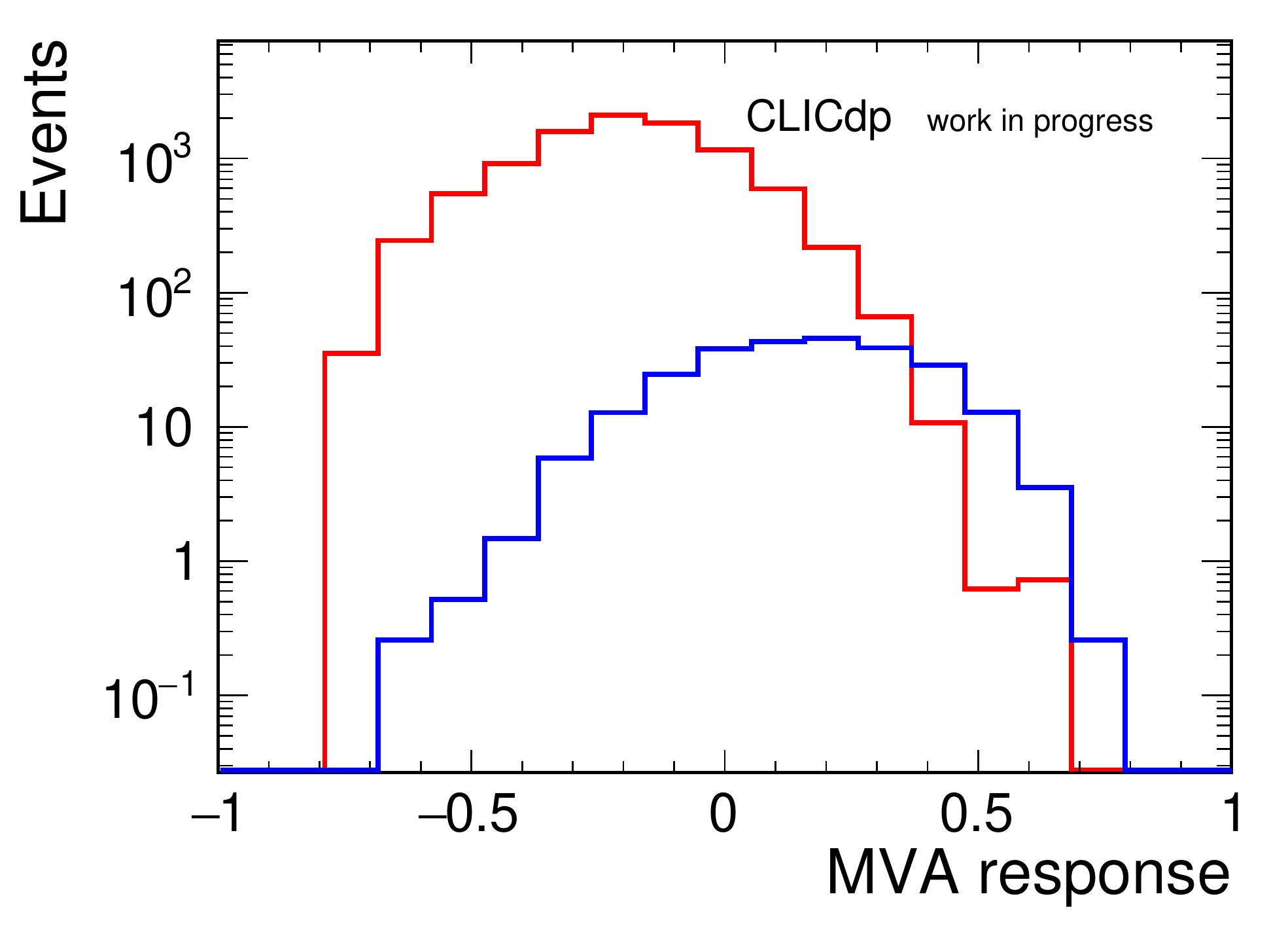}
\includegraphics[width=0.49\textwidth]{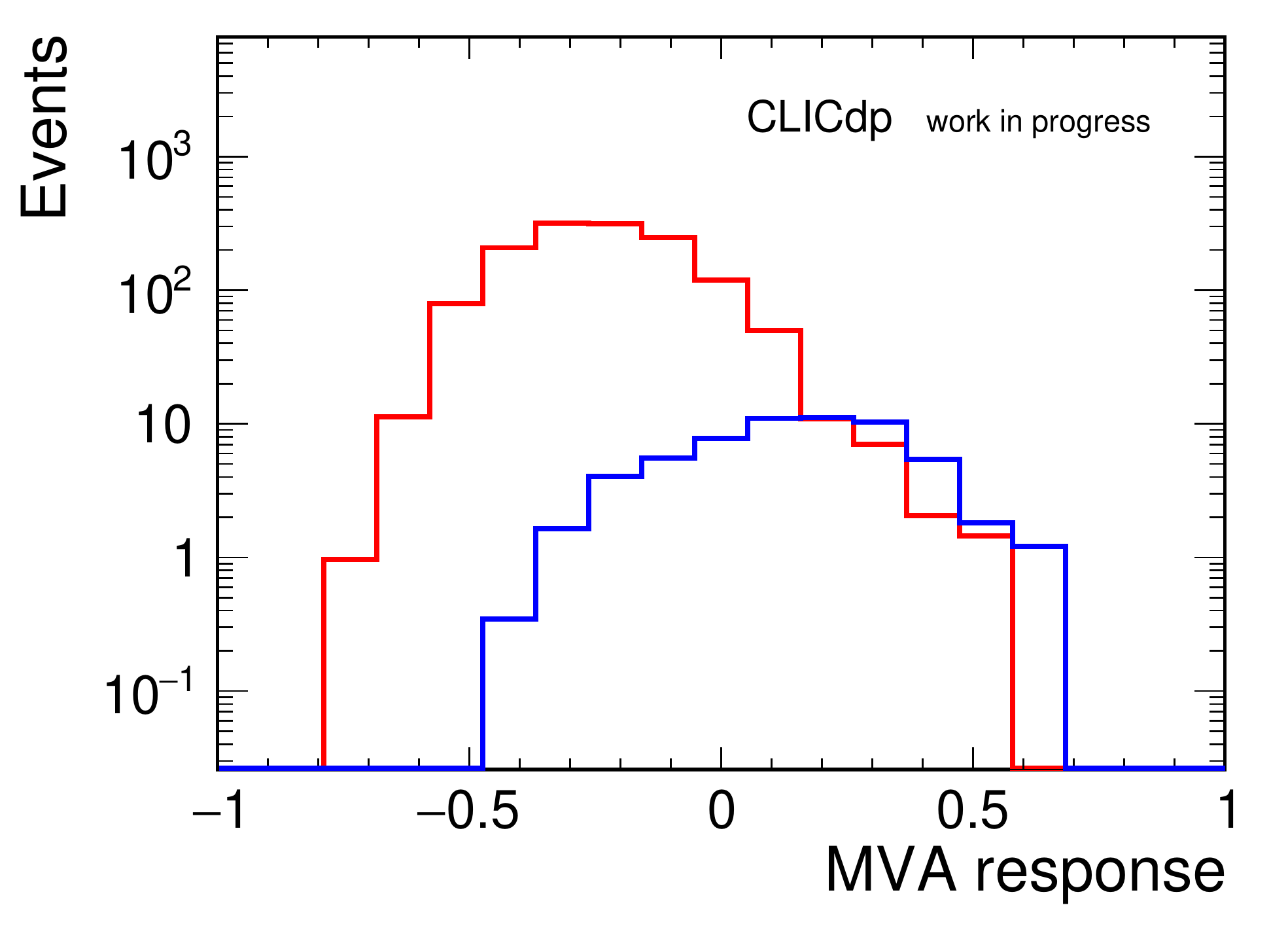}
\end{center}
\caption{ Response distribution of the BDT classifiers used for the final
    signal event selection for hadronic (left) and semi-leptonic
    (right) events. The background (red histogram) sample is
    normalized to 500~fb$^{-1}$ while the 
    signal (events with FCNC decay; blue histogram)
    to \mbox{BR(\tch)}$\times$\mbox{BR(\hbb)} $=10^{-3}$ for the same
    integrated luminosity.
\label{fig:tch_bdt}  }
\end{figure}

In the final step of the analysis, kinematic fits are performed for
signal and background hypotheses.
The jet combination which minimises the $\chi^2$ value for the event is
selected separately for each hypothesis.
Flavour tagging results are used to reduce the number of possible
configurations.
Results from kinematic fits and the flavour tagging results are
then used as an input for the final BDT selection optimised to discriminate
between signal and background events.
Resulting response distributions from the BDT classifiers trained to
select signal events in hadronic and semi-leptonic samples are
presented in Fig.~\ref{fig:tch_bdt}.
%
%
Final cuts on the BDT response were optimised to obtain the best
expected limit on FCNC branching ratio, for the assumed luminosity of
500~fb$^{-1}$.
The resulting cuts turned out to be relatively tight.
The final selection efficiency for FCNC events is 10.4\% while the
background suppression is at the level of $10^{-4}$.
This results in the expected 95\% C.L. limits of
\begin{eqnarray}
  \textrm{BR}(\mtch) \times \textrm{BR}(\mhbb)  & < & 1.6 \cdot 10^{-4}
\end{eqnarray}   
for 500 fb$^{-1}$ collected at 380 GeV (hadronic and semi-leptonic
samples combined).
Details of the event selection efficiency for hadronic and
semi-leptonic channel are presented in Tab.~\ref{tab:tch_effi}.
\begin{table}
  \begin{center}
\begin{tabular}{l|c|c|c|c}
  & \multicolumn{2}{c|}{Hadronic}
  & \multicolumn{2}{c}{Semi-leptonic} \\ \cline{2-5}
  & Signal & SM $\PQt\PAQt$ 
  & Signal & SM $\PQt\PAQt$ \\ \hline
  Classification & 0.66 & 0.42 & 0.19 & 0.28 \\  
  Flavour tagging & 0.54 & 0.059 & 0.42 & 0.013 \\
  Event quality & 0.89  & 0.90  & 0.92  &  0.90  \\
  Final MVA cut & 0.23 & 0.0038 & 0.44 & 0.013 \\ \hline
  Total  & 0.072 & 0.000086 & 0.032 & 0.000044
 \end{tabular}
  \end{center}
  \caption{Summary of selection efficiency for analysis of the FCNC top
    quark decay \tch. Shown are the selected fractions of
    hadronic and semi-leptonic events, for FCNC signal and SM top pair
    production background, at different analysis stages as well as the
    total selection efficiency.  
    \label{tab:tch_effi}}
\end{table}

\section{Conclusions}

Considered in this contribution was  the feasibility of measuring the FCNC
top decay \tcg and \mbox{\tch} at CLIC running at 380 GeV.
Results based on the full detector simulation were presented, taking
luminosity distribution, beam polarization and beam induced background
into account. 
For the \tcg decay, based on the analysis of the hadronic channel
only, the expected sensitivity (expected 95\% C.L. limit on the FCNC
branching ratio) is $3 \cdot 10^{-5}$, while for the \tch decay the
corresponding limit is $1.6 \cdot 10^{-4}$ (including Higgs branching
ratio to $\PQb\PAQb$). 
Presented results should be considered as a ``Work in Progress'' report,
the study is ongoing and publication of the final results is in preparation.

\section*{Acknowledgments}

This work benefited from services provided by the ILC Virtual
Organisation, supported by the national resource providers of the EGI
Federation.



\printbibliography[title=References]

\end{document}